\documentclass[aps,pra,preprint,superscriptaddress,amsmath,amssymb,floatfix]{revtex4}

\usepackage{times}

\usepackage{graphicx,amsmath,amssymb,amsfonts}
\usepackage{textcomp,url}
\usepackage{bm}
\usepackage{rotating,color}
\newcommand{\kb}{{\bf k}}

\begin{document} 

\author{Benjamin J. Galow}
\affiliation{Max-Planck-Institut f\"{u}r Kernphysik, Saupfercheckweg 1, 
D-69029 Heidelberg, Germany}
\author{Zolt\'an Harman*}
\affiliation{Max-Planck-Institut f\"{u}r Kernphysik, Saupfercheckweg 1, 
D-69029 Heidelberg, Germany}
\affiliation{ExtreMe Matter Institute EMMI, Planckstrasse 1,
64291 Darmstadt, Germany}
\author{Christoph H. Keitel}
\affiliation{Max-Planck-Institut f\"{u}r Kernphysik, Saupfercheckweg 1,
D-69029 Heidelberg, Germany}

%\date{\today}

%\pacs{52.38.Kd, 37.10.Vz, 42.65.-k, 52.75.Di, 52.59.Bi, 52.59.Fn, 41.75.Jv, 87.56.bd}

\title{Intense high-quality medical proton beams via laser fields}

\begin{abstract}
During the past decade, the interaction of high-intensity lasers with solid targets  
has attracted much interest, regarding its potential in accelerating charged particles.
 In spite of tremendous progress in laser-plasma based acceleration, it is still not
clear which particle beam quality will be accessible within the upcoming multi petawatt (1 PW = 10$^{15}$ W) laser generation.
Here, we show with simulations based on the coupled relativistic equations of motion that protons stemming from 
laser-plasma processes can be efficiently post-accelerated using crossed laser beams focused to 
spot radii of a few laser wavelengths. We demonstrate that the crossed beams produce 
monoenergetic accelerated protons with kinetic energies $> 200$ MeV, small energy spreads ($\approx$ 1$\%$)
and high densities as required for hadron cancer therapy.
To our knowledge, this is the first scheme allowing for this important
application based on an all-optical set-up. 
\end{abstract}

\maketitle

Accelerating charged particles is of paramount importance in a wide variety of fields, ranging from medicine \cite{med1,med2,cancer,debus1} and 
material science \cite{lithog} to being used to resolve the smallest structures of our universe \cite{lhc,ledingham}.
The rapid development of high-intensity laser systems which are likely
to exceed in the near future 10$^{25}$ W cm$^{-2}$~\cite{eli,hiper} rendered particle beam creation by laser-matter interaction 
feasible. Laser-driven accelerators offer the unique feature of ultra-high electric field gradients of several TV 
m$^{-1}$, outperforming those in conventional accelerators by more than six orders of magnitude and thus offering the 
possibility of compact and low-cost devices~\cite{dunne-pers}. Electron~\cite{plasma-el} and proton beams have recently 
been generated by focusing high-intensity laser light onto solid targets. This mechanism of target normal sheath 
acceleration (TNSA)~\cite{plasma1,plasma2,plasma3,plasma4,plasma5,plasma6,plasma7,plasma8,plasma9,plasma10,plasma11,willi} is 
realized by the strong quasi-static electric field induced by the ionization and acceleration of electrons by the intense 
laser field. A further laser-plasma-interaction process, the skin-layer ponderomotive acceleration (S-LPA) resulting from 
the huge electric potential gradient of the plasma leads to ion beams of high density~\cite{badziak}. To date the energy 
resolution and the ion kinetic energy of the generated beams merely reach the parameters required for skin-deep neoplasm, 
i.e. excluding deep-seated tumors. Furthermore there is still controversy~\cite{plasma4} which beam quality, i.e. which 
total particle number, kinetic energy and energy spread will be accessible with the forthcoming multi petawatt-class laser 
systems~\cite{eli,hiper}.

\begin{figure}[ht]
\footnotesize
\includegraphics[width=69mm]{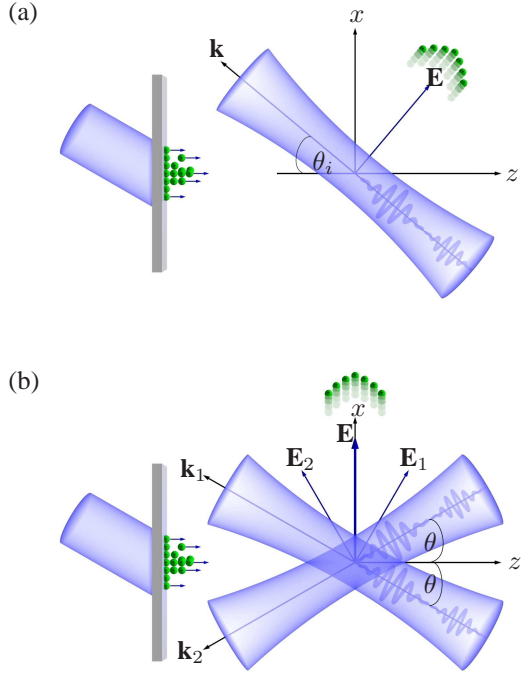}
\begin{picture}(0,0)(+137,11)
\put(49.5,140){$x$}
\put(109,82){$z$}
\put(44,130){$\textbf{E}$}
\put(-15,117){$\textbf{k}_1$}
\put(-16,48){$\textbf{k}_2$}
\put(77,89){$\theta$}
\put(77,73){$\theta$}
\put(68,121){$\textbf{E}_1$}
\put(25,121){$\textbf{E}_2$}
\put(49,289){$x$}
\put(108,229){$z$}
\put(78,265){$\textbf{E}$}
\put(-4,276){$\textbf{k}$}
\put(35,233){$\theta_i$}
\put(-80,290){(a)}
\put(-80,150){(b)}
\end{picture}
\vspace{-12mm}
\normalsize
\caption{Laser-proton acceleration schemes. (a)
The protons, produced by laser-plasma interaction, are injected with the angle $\theta_i$ with respect to the propagation
direction of a pulsed beam through its focus. The laser field polarization is denoted by ${\bf E}$ and the propagation direction 
is given by $\kb$. The protons are ejected out of the focus in the polarization direction ${\bf E}$. (b) Here, the protons 
are injected through the intersection point of two pulsed beams with crossing half-angle $\theta$. The laser field 
polarizations are denoted by ${\bf E}_1$ and ${\bf E}_2$ and their propagation directions are given by $\kb_1$ and 
$\kb_2$. The protons are ejected in direction of the resulting electric field ${\bf E}$.
}
\label{fig1}
\end{figure}

\begin{figure}[t]
\footnotesize
\includegraphics[width=73mm]{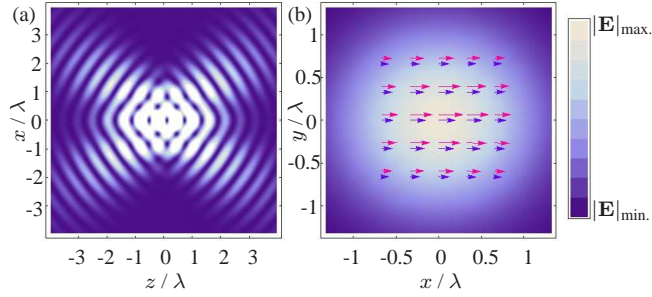}
\begin{picture}(0,0)(15,15)
\begin{scriptsize}  
\put(-207,98){{(a)}}
\put(-200,90){3} 
\put(-200,79){2}
 \put(-200,68){1} 
\put(-200,57){0} 
\put(-202,47){-1} 
\put(-202,36){-2} 
\put(-202,24){-3}

\put(-186,6){-3} 
\put(-175,6){-2}
\put(-164,6){-1} 
\put(-151,6){0} 
\put(-140,6){1} 
\put(-129,6){2} 
\put(-118,6){3}
\put(-208,52){\begin{sideways}$x$ / $\lambda$\end{sideways}} 
\put(-157,-3){$z$ / $\lambda$} 
\put(-103,98){{(b)}}
\put(-95,90){1} 
\put(-100,74){0.5} 
\put(-96,57){0} 
\put(-103,42){-0.5}
\put(-98,25){-1}

\put(-82,6){-1} 
\put(-68,6){-0.5}
\put(-47,6){0}
\put(-33,6){0.5}
\put(-14,6){1}

\put(-103,52){\begin{sideways}$y$ / $\lambda$\end{sideways}} 
\put(-53,-3){$x$ / $\lambda$} 

\put(12,94){$|\bf E|_{\text{\tiny max.}}$}
\put(12,24){$|{\bf E}|_{\text{\tiny min.}}$}
%\put(12,47){\begin{sideways}{\footnotesize Electric field strength}\end{sideways}}
%\put(){}
\end{scriptsize}
\end{picture}
\vskip 5mm
\normalsize
\caption{Laser intensity and field strength.
(a) Intensity profile of two crossed Gaussian beams in the propagation plane $y=0$. For visualization purpose the 
crossing-half angle is set to $\theta=35^\circ$. The brighter areas correspond to higher field intensity. (b) Vector field 
plot of the polarization plane $z=0$ at $t=0.2$ fs for the single beam scheme (blue arrows) with power $P$ and crossed 
beam scheme (red arrows) with power $P/2$ for each beam. The constructive interference of the crossed beams results in a 
higher electric field strength in the intersection volume. The background of the graph shows a density map of the electric 
field strength $|{\bf E}|$ of two crossed beams in the polarization plane.}
\label{fig2}
\end{figure}

In this article we investigate the vacuum post-acceleration of plasma-generated ions by means of 
lasers in a single or crossed-beams configuration. 
This scheme of laser particle acceleration was first proposed for electrons~\cite{haaland,esarey1,sal00}. The basic idea 
is to send the protons originating from a TNSA or S-LPA experiment through the crossing point of two laser beams at a 
half-angle $\theta$ with respect to the $z$-axis (see Fig.~1b for a scheme and a coordinate system). Within the simple 
plane-wave picture and assuming the laser fields to have the same amplitude, frequency, phase and to be polarized as shown 
in Fig.~1b, the resultant electric field component along the symmetry axis of the set-up vanishes for all points on the 
$x$-axis. At the same time, the $x$-component, increased by constructive interference, violently accelerates the 
particles. Such a coherent combination of intense beams is experimentally feasible (see p. 52 of Ref.~\cite{eli}).
Subsequent motion of the charged particle ejected from the focal region and no longer interacting with the 
laser pulse may be taken as linear. The results of our theoretical simulations demonstrate that the protons gain kinetic 
energies larger than $200$ MeV (employing two crossed beams each with a peak intensity of
1.9$\times$10${}^{24}$ W cm$^{-2}$) with an energy spread of roughly 1$\%$.
Furthermore (assuming a realistic repetition rate of 10 Hz~\cite{pfs,eli}), the total number of generated protons
reaches 10$^{10}$ min$^{-1}$. For the first time, all requirements are fulfilled for broader radio-oncological
use~\cite{med1,med2} based on an optical accelerator.\\
\mbox{}\\
{\bf\Large Results}\\
\mbox{}\\
In order to generate ultra-strong accelerating fields~\cite{eli,hiper} of 10$^{24}$ W cm$^{-2}$, one needs to focus the 
laser field to a beam waist radius on the order of the laser wavelength~\cite{bahk}. This necessitates an accurate description of the 
fields beyond the widely-used paraxial approximation. The parameters of a linearly polarized Gaussian beam which 
propagates in the $z$-direction and is polarized in the $x$-direction and subsequently rotated by $\theta_i$ 
(Fig.~\ref{fig1}a) will be used to model the fields, i.e. the beam waist radius $w_0$, the Rayleigh length $z_r=\pi 
w_0^2/\lambda$, where $\lambda$ is the laser's wavelength, and the diffraction angle is $\varepsilon=w_0/z_r=\lambda/(\pi 
w_0)$. The expressions giving the Cartesian field components $E_x$, $E_y$, $E_z$, $B_x$, $B_y$, $B_z$, as well as the 
expression for the power of the fields to order $\varepsilon^{11}$ in the diffraction angle can be found in 
the Methods section and for more details we refer to
Refs.~\cite{sal-apb,sal-prl2}. For the intensity profile of the employed Gaussian beams see Fig.~\ref{fig2}.

High-intensity laser systems provide their energy in short pulses which are already sufficient to accelerate particles to 
high velocities~\cite{scheid}. Employing pulsed fields also ensures that the particles injected into the focus get captured
rather than reflected. To the lowest order in time, this can be described by multiplying the electromagnetic field 
components with a Gaussian temporal envelope factor,
\begin{eqnarray}
\bm{E}&\to& \exp\left(-\frac{(t-z/c)^2}{2\Delta t^2}\right)\bm{E},\nonumber\\
\bm{B}&\to& \exp\left(-\frac{(t-z/c)^2}{2\Delta t^2}\right)\bm{B},
\end{eqnarray}
with $\Delta t$ defined via the Full Width at Half Maximum (FWHM) pulse duration $\Delta t_{\text{\tiny 
FWHM}}=2\sqrt{2\log{2}}\Delta t$. This approximation is valid for $T/\Delta t_{\text{\tiny FWHM}}\ll1$, with $T$ being the 
laser period. For the titanium-sapphire laser with wavelength $\lambda=0.8~\mu$m ($T=2.65$ fs) and pulse durations 
of $\Delta t\gtrsim 10$ fs used in our simulations, this turns out to be an adequate description. Hence, further temporal 
corrections~\cite{space-temp} which describe the field solutions as a dual power series in the diffraction 
angle $\varepsilon$ and in the small ratio $T/(2\pi\Delta t)$ can be neglected.

The motion of an ensemble of $N$ identical particles of mass $m$ and charge $q$ in the electric and magnetic fields $\bm{E}$ 
and $\bm{B}$, respectively, of a laser beam is considered classically, with randomized initial distributions.
The use of laser systems of high intensity (exceeding 
10$^{24}$ W cm$^{-2}$ for protons) requires a relativistic treatment of particle motion. Thus, the dynamics is governed by 
the coupled Newton-Lorentz (or energy-momentum transfer) equations (given in SI units):
\begin{eqnarray}\label{motion_coul}
\frac{d\bm{p}_j}{dt} &=& q \left( \bm{E}(\bm{r}_j)+\bm{E}_{j}^{\text{\tiny{int.}}}+c\bm{\beta}_j\times\left(\bm{B}
(\bm{r}_j)+\bm{B}_{j}^{\text{\tiny{int.}}}\right)
 \right), \nonumber \\
\frac{d{\cal{E}}_j}{dt} &=& qc\bm{\beta}_j\cdot \left( \bm{E}(\bm{r}_j)+
\bm{E}_{j}^{\text{\tiny{int.}}} \right).  \nonumber\\
\end{eqnarray}
The relativistic energy and momentum of a given particle labeled with $j$ are denoted here by ${\cal E}_j=\gamma_j mc^2$ 
and $\bm{p}_j=\gamma_j mc\bm{\beta}_j$, respectively, with $\bm{\beta}_j=\bm{v}_j/c$ its velocity scaled by $c$, and 
$\gamma_j=(1-\beta_j^2)^{-1/2}$ its Lorentz factor. The fields mediating inter-ionic interaction are 
given in the Methods section.
To obtain the kinetic energy gained by interaction with a laser beam, numerical solutions of the equations of motions will 
be sought. A numerical integration of equation (\ref{motion_coul}) yields $\bm{\beta}_j^{\rm fin}$ and, hence, $\gamma_j^{\rm 
fin}$ at a later final time $t^{\rm fin}$ taken equal to many laser field cycles. Finally, one calculates the final 
kinetic energy of the particle from $K_j^{\rm fin}=\gamma_j^{\rm fin} mc^2$.

Foremost, we carry out simulations based on the coupled equations of motion equation (\ref{motion_coul}) for an ensemble of 50 
particles at later used particle densities in order to determine the dominant nearest-neighbor
contribution of proton-proton interaction effects on 
the resulting particle beam. Due to the dominating ponderomotive laser forces which lead to a fast drifting apart of the 
ensemble's particles, for relativistic laser intensities, i.e. $>10^{24}$ W/cm$^2$ for protons, it turns out that the 
dynamics, the energy gain and its spread are influenced negligibly by inter-ionic interaction. This is in contrast to the 
non-relativistic laser regime. Here, the repulsive Coulomb interaction is the prevailing part of the 
interaction, which is in this case non-negligible compared to the electromagnetic fields of the laser. As a consequence, 
the accelerated ions occupy a larger phase space volume. From radio-oncological point of view we are interested in proton 
beams of fully relativistic energies, hence for further calculations at these energies and densities it is sufficient to 
study the uncoupled equations of motions only, i.e. we set 
$\bm{B}_{j}^{\text{\tiny{int.}}}=\bm{E}_{j}^{\text{\tiny{int.}}}=\bm{0}$ in equation (\ref{motion_coul}).

The definition of a coordinate system for the crossed beams set-up depicted in Fig.~\ref{fig1}b is given by the coordinate 
transformations
        $x_1 = x\cos\theta-z\sin\theta$, $y_1=y$,
        $z_1=x\sin\theta+z\cos\theta$ for the first beam and
        $x_2 = x\cos\theta+z\sin\theta$, $y_2=y$,
        $z_2=-x\sin\theta+z\cos\theta$
for the second beam, respectively.
The resulting field components which enter equation (\ref{motion_coul}) are
\begin{eqnarray}
        E_x &=&
        (E_{1x}+E_{2x})\cos\theta+(E_{1z}-E_{2z})\sin\theta,\nonumber\\
        E_y &=& E_{1y}+E_{2y},\nonumber\\
        E_z &=&
        (-E_{1x}+E_{2x})\sin\theta+(E_{1z}+E_{2z})\cos\theta\nonumber , \\
        B_x &=&(B_{1z}-B_{2z})\sin\theta,\nonumber\\
        B_y &=& B_{1y}+B_{2y},\nonumber\\
        B_z &=&(B_{1z}+B_{2z})\cos\theta.
\label{res_fields}
\end{eqnarray}
Choosing a small crossing half-angle $\theta$ leads to constructive addition of the dominating $x$-components of the 
electric fields in equation (\ref{res_fields}). The adding of further laser beams would further increase the laser intensity 
and hence the exit kinetic energy of the accelerated particles. With respect to the energy spread one cannot achieve 
substantial improvement, however. For all subsequent simulations we restrict our analysis to the case of two crossed beams 
and set $\theta=3^\circ$.

\begin{figure}[b]
\footnotesize
\includegraphics[width=70mm]{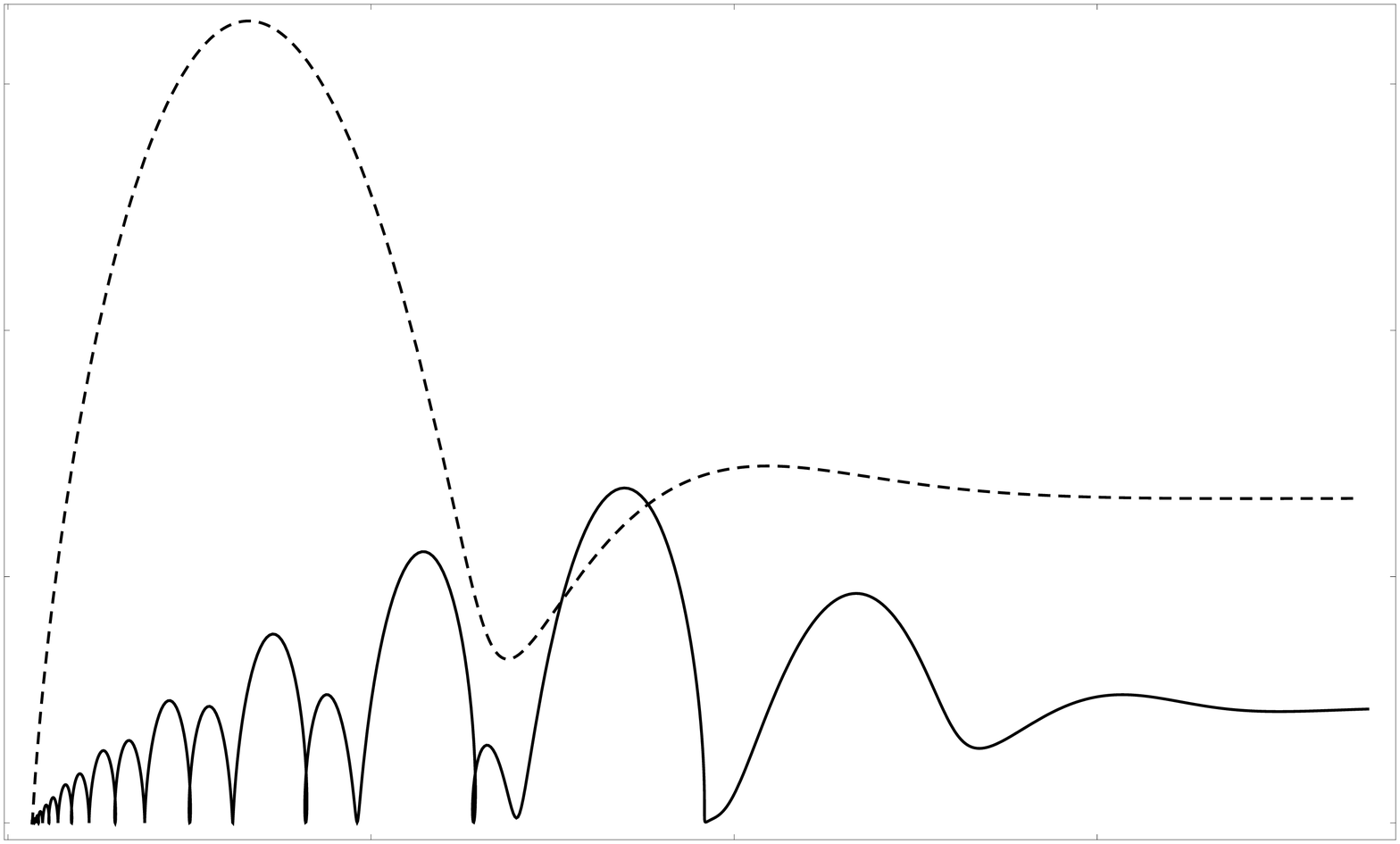}
\begin{picture}(0,0)
\put(-207,-3){0}
\put(-215,37){0.5}
\put(-215,72){1.0}
\put(-215,107){1.5}

\put(-154,-5){0.5}
\put(-102,-5){1.0}
\put(-50.5,-5){1.5}

\put(-105,-17){$z/\lambda$}
\put(-223,50){\begin{sideways}$K$ [GeV]\end{sideways}}
\end{picture}
\vskip 5mm
\label{traject}
\normalsize
\caption{Proton dynamics. Kinetic energy $K$ for one proton initially being at rest and located at $x=z=\lambda/30$ and $y=0$. The initial 
time is $t_i=0$ (dashed line) and $t_i=-10\Delta t$ (full line). The single proton dynamics is compared at same laser 
parameters.}
\end{figure}

In order to simulate a realistic particle injection into the focal point of the laser pulses, we consider an ensemble of 
5000 particles initially randomly distributed in a micron-scale volume $V_{\text{\tiny{focus}}}$ oriented along the 
$z$-axis. Initially, we have to ensure that the ensemble is not already exposed to the laser fields. This is realized by 
starting the simulations at time $t\leq-5\Delta t$. At these initial times the electromagnetic fields are damped by the 
Gaussian envelope factor such that the particles' motion is only negligibly influenced by the external laser fields. 
Fig.~3 compares the motion of a proton starting at the same spatial point but having two different initial times. One can 
see that in the unphysical case when the particle is directly exposed to the laser fields (dashed line), it immediately 
gains energy and is ejected out of the focus after one laser cycle. In a realistic setting, we choose an initial time of 
$t=-10\Delta t$: the particle slowly starts to oscillate and then gets captured and accelerated by the approaching pulse 
(full line). Choosing the initial time to be $t=0$ would lead to an energy gain overestimation by a factor of 
approximately three.

\begin{table}[b]
\caption{Variation of laser parameters.
Average particle kinetic energy $\bar{K}$ and its percentual spread for different laser system parameters. $N_i=n_i\cdot 
V_{\text{\tiny{focus}}}$ is the number of ions one can accelerate as one bunch with $n_i^{\text{\tiny{S-LPA}}}
\approx10^{21}\rm{~cm}^{-3}$ and $n_i^{\text{\tiny{TNSA}}}\approx10^{19}\rm{~cm}^{-3}$ is the ion density of the source used and 
$V_{\text{\tiny{focus}}}$ denotes the volume initially containing all ions. The crossing half-angle is $\theta=3^\circ$. 
The optimal particle injection angle for the single beam set-up is $\theta_i=3^\circ$ for the S-LPA source and 
$\theta_i=50^\circ$ in case of the TNSA source, respectively. For two crossed beams the particles are injected with an 
angle $\theta_c$ with respect to the symmetry axis ($z$-axis) of the laser beam configuration. In case of the S-LPA source 
we have $\theta_c=0^\circ$ and for the TNSA source $\theta_c=50^\circ$.}
\begin{ruledtabular}
\begin{tabular}{ccc}
~        &     $\bar{K}$ [MeV]  & $N_i$   \\

\hline
\multicolumn{3}{c}{S-LPA source, $P$=10 PW, $\Delta t$=19.2 fs, $w_0=1\lambda$}\\
single &      28.1  $\pm$ 1.2 \%  &  1.0$\cdot$10$^6$   \\
crossed &      59.4 $\pm$  1.0 \%       &  1.0$\cdot$10$^6$      \\

\hline
\multicolumn{3}{c}{S-LPA source, $P$=40 PW, $\Delta t$=10.7 fs, $w_0=1\lambda$}\\
single &       113.2 $\pm$ 1.6 \% &     1.0$\cdot$10$^6$ \\
crossed &      233 $\pm$ 1.0 \% &       1.0$\cdot$10$^6$ \\

\hline
\multicolumn{3}{c}{S-LPA source, $P$=100 PW, $\Delta t$=23.8 fs, $w_0=2\lambda$}\\
single &       73.2 $\pm$ 1.6 \% &  1.3$\cdot$10$^7$   \\
crossed &      152  $\pm$ 1.0 \% &       1.3$\cdot$10$^7$ \\

\hline
\multicolumn{3}{c}{TNSA source, $P$=100 PW, $\Delta t$=14.4 fs, $w_0=2\lambda$}\\
%\hline
single&       64.6 $\pm$ 0.7 \% & 1.0$\cdot$10$^5$    \\
crossed &      141  $\pm$ 0.5 \% &       1.0$\cdot$10$^5$ \\

\end{tabular}
\end{ruledtabular}
\end{table}

The volume initially containing the particle ensemble has a length of the order of the laser wavelength and a radius of 
tens of nanometers, dependent on the focus diameter of the applied laser system, ensuring that all protons are exposed to 
a homogeneous field. The particles will be assumed to possess initial kinetic energies distributed normally around a mean 
value $\bar{K}$ and having a spread $\Delta K$. As a source we take protons originating from laser-plasma interactions, 
such as the S-LPA mechanism, with $\bar{K}=17$ keV~\cite{badziak} and assuming a large energy spread of $\Delta K=100$\,\% 
or from the TNSA mechanism, with $\bar{K}=1.2$ MeV and $\Delta K=25$\,\%~\cite{plasma2}. The total power and the beam 
pulse duration are varied. Note that the peak intensity of one linearly polarized 10 PW laser beam focused to 
$w_0=1\lambda$ is already $I\sim9.6\times10^{23}$ W cm$^{-2}$~\cite{eli,hiper}.\\
\mbox{}\\
{\bf\Large Discussion}\\
\mbox{}\\
Our main interest is in the energy gain, or exit kinetic energy, of the nuclei, their trajectories and, hence, the aspects 
that determine the quality of an accelerated beam of such nuclei. In Tab.~1, simulation results for the laser acceleration 
are summarized. The single and crossed beams scheme are compared at same total laser power $P$, pulse duration $\Delta t$ 
and focus radius $w_0$. The energy gain is ranging from 59 MeV to 233 MeV in case of the crossed beams setup and from 28 
MeV to 113 MeV for injection of the ensemble into the focus of only one beam, respectively. Looking at the energy spread 
one can see that for the crossed beams it is always $1\%$ and a little higher for the case of the single beam setup.

Moreover, one can see from Tab.~1 that for the systems studied here, the average exit kinetic energy $\bar{K}$ obeys for 
constant injection angle and injection energy the rough scaling behavior $\bar{K}\propto I\propto P/w_0^2$. This behavior 
results from the fact that the optimal acceleration regime depends strongly on the pulse duration $\Delta t$ rather than 
only on the electric field strength. It is achieved for the laser-particle interaction length being of the same order of 
the Rayleigh length. In Ref.~\cite{zolotorev} the same scaling behavior was derived for electrons. Therefore, in order to 
maximize the energy gain of the accelerated protons we first choose the laser power $P$ and the focus radius $w_0$, and 
then adjust to the optimal pulse duration $\Delta t$.

A further scaling law can be derived for the particle number, which is proportional to the focal volume. Using 
$V_{\text{\tiny{focus}}}\propto w_0^2\cdot z_r$ and the definition of the Rayleigh length $z_r\propto w_0^2$, one obtains 
$V_{\text{\tiny{focus}}}\propto w_0^4$. The typical particle number needed for ion cancer treatment is 
$10^6$-$10^{10}$ per shot with a repetition rate on the order of 5 Hz~\cite{eickhoff}, depending on the ionic species. For 
a typical interaction volume, one needs at least an ion density of 
$n_i=10^6/V_{\text{\tiny{focus}}}\approx10^{20}\rm{\,cm}^{-3}$ in order to render our scheme feasible for medical 
applications. Using plasma-generated protons one obtains with the TNSA mechanism an ion density up to the order of 
$n_i^{\text{\tiny{TNSA}}}\approx10^{19}\rm{\,cm}^{-3}$ and the S-LPA ion source generates a density of 
$n_i^{\text{\tiny{S-LPA}}}\approx10^{21}\rm{\,cm}^{-3}$~\cite{badziak}. The latter yields particle numbers of $10^7$ per 
laser shot. Combined with lasers operated at 10 Hz repetition rate~\cite{eli,pfs} this ion number is sufficient for cancer 
therapy while for the TNSA mechanism a modest improvement would still be necessary.

Such improvement can e.g. be achieved by substituting the assumed titanium-sapphire laser by a super-intense CO$_2$ laser 
with a typical wave length of $\lambda=10.6\,\mu\rm{m}$. Using the fact that the waist radius $w_0\propto\lambda$ and 
hence the Rayleigh length $z_r\propto \lambda$, the focal region thus increases by three orders of magnitude. 
Consequently, the ion density needed decreases by three orders of magnitude and the needed laser intensity by two orders 
of magnitude. In the near future the use of high repetition rate laser systems~\cite{eli,pfs} will further decrease the 
number of protons needed per shot.

The combination of the laser-plasma mechanism as a proton source and the post acceleration process by means of single or crossed 
laser beams places laser acceleration of particles on the cusp of medical feasibility utilizing present-day or near-future 
laser technology~\cite{eli,hiper,pfs}, while anticipated costs are presently on the scale of those for current synchrotron facilities
(cf. \cite{hit} and \cite{eli-costs}).
The rapid advancement of laser technology renders significant reduction likely for the near future.
All requirements needed for broader radio-oncological use may be achieved: sufficient proton density and exit kinetic energies, and sharp energy spread
of approximately 1\,\%.
The scheme that we introduce in the present work calls for tight 
focusing mechanisms \cite{bahk} and relies on results of laser-plasma-interaction
research~\cite{plasma1,plasma2,plasma3,plasma4,plasma5,plasma6,plasma7,plasma8,plasma9,plasma10,plasma11,willi}.
\\
\mbox{}\\
{\bf\Large Methods}\\
\mbox{}\\
{\bf Description of a tightly focused linearly polarized beam}\\
\mbox{}\\
The electromagnetical field components of the tightly focused
linearly polarized beams which we employ
in our simulations can be given as a power series in the diffraction
angle $\varepsilon$. 
Their functional form is given by
\begin{eqnarray}
 E_x &=& E\left\{S_1
  	+\varepsilon^2\left[\xi^2S_3-\frac{\rho^4S_4}{4}\right]
  	+\cdots+{\cal O}(\varepsilon^{10})\right\},\nonumber\\
   E_y &=& E\xi\upsilon\left\{\varepsilon^2\left[S_3\right]
  	+\cdots+{\cal O}(\varepsilon^{10})\right\},\nonumber\\
   E_z &=& E\xi\left\{\varepsilon\left[C_2\right]
  	+\cdots+{\cal O}(\varepsilon^{11})\right\},\nonumber\\
   B_x &=& 0, \nonumber\\
   B_y &=& \frac{E}{c}\left\{S_1+\varepsilon^2\left[\frac{\rho^2S_3}{2}-\frac{\rho^4S_4}{4}\right]
  	+\cdots+{\cal O}(\varepsilon^{10})\right\},\nonumber\\
   B_z &=& \frac{E}{c} \upsilon\left\{\varepsilon\left[C_2\right]
  	+\cdots+{\cal O}(\varepsilon^{11})\right\}.\label{field-detail}
\end{eqnarray}
With $\omega$ the angular frequency of the fields, $\xi=x/w_0$, 
$\upsilon=y/w_0$, $\zeta=z/z_r$, $\psi_G=\tan^{-1}\zeta$, and 
$r=\sqrt{x^2+y^2}$, $\rho=r/w_0$, the remaining symbols in equation
(\ref{field-detail}) have the following definitions
\begin{eqnarray}
  \label{E} E &=& E_0 e^{-r^2/w^2};\quad w = w_0\sqrt{1+\zeta^2},\nonumber \\
  \label{Cn} C_n &=& \left(\frac{w_0}{w}\right)^n \cos(\psi+n\psi_G);\quad n=1, 2, 3, \cdots, \nonumber\\
  \label{Sn} S_n &=& \left(\frac{w_0}{w}\right)^n \sin(\psi+n\psi_G),
\end{eqnarray}
where
\begin{equation}
  \label{ps} \psi = \psi_0+\omega t-kz-\frac{kr^2}{2R};\quad R =
  z+\frac{z_r^2}{z},
\end{equation}
and $\psi_0$ is a constant initial phase. Also, $t$ is the time 
and $k=2\pi/\lambda$ is the wavenumber. On the other hand, with $E_0\to 
E_{0l}$, the power expression may be given, to the same order in 
$\varepsilon$ as the field components, by
\begin{eqnarray}\label{pow-linear}
  \label{pow} P_l &=& \frac{\pi w_0^2}{4}\frac{E_{0l}^2}{c\mu_0}
  \left[1 + \left(\frac{\varepsilon}{2}\right)^2 + 2\
\left(\frac{\varepsilon}{2}\right)^4 + 6\
\left(\frac{\varepsilon}{2}\right)^6 \right.\nonumber\\
  & &\left.+ \frac{45}{2}
\left(\frac{\varepsilon}{2}\right)^8 + \frac{195}{2}
\left(\frac{\varepsilon}{2}\right)^{10} \right],
\end{eqnarray}
where $c$ is the speed of light in vacuum, $\mu_0$ is the permeability 
of free space and $E_{0l}$ is the electric field amplitude, with $l$ 
standing for linearly polarized. Note that 
$E_{0l}\propto\sqrt{P_l}$ and that the leading term in $E_{0l}$ is 
inversely proportional to $w_0$. For more details we refer to Refs.~\cite{sal-apb,sal-prl2}.\\
\mbox{}\\
{\bf Relativistic coupled equations of motion}\\
\mbox{}\\
The fields mediating the inter-ionic interaction in the relativistic
coupled equations of motion (\ref{motion_coul}) are modeled by 
$\bm{E}_{j}^{\text{\tiny{int.}}} = \sum_{k \ne j}\left({-\nabla\phi_{jk}-\frac{\partial}{\partial t}\bm{A}_{jk}}\right)$ 
and $\bm{B}_{j}^{\text{\tiny{int.}}}=\sum_{k \ne j}\left(\nabla\times\bm{A}_{jk}\right)$ with $j,k\in \{1,2,\dots,N\}$. 
The interaction potentials read
\begin{eqnarray}
\phi_{jk}&=&\frac{q}{4\pi\epsilon_0}\frac{1}{|\bm{r}_j-\bm{r}_k|}\label{Coulomb} \,,\\
\bm{A}_{jk}&=&\frac{q}{8\pi\epsilon_0 c^2 |\bm{r}_{jk}|}\left(\bm{v}_k +
\frac{\bm{r}_{jk}(\bm{v}_k\bm{\cdot}\bm{r}_{jk})}{|\bm{r}_{jk}|}\right)\label{Darwin} \,,
\end{eqnarray}
with the relative displacement $\bm{r}_{jk}=\bm{r}_j-\bm{r}_k$ and $\epsilon_0$ being the vacuum permittivity. 
Equation (\ref{Coulomb}) is the scalar part of the interaction given by the Coulomb potential, whereas relativistic effects 
such as retardation and current-current interaction are included in the Darwin vector potential up to 
$\mathcal{O}(\beta^2)$~\cite{jackson} in equation (\ref{Darwin}). Typical kinetic energies of the accelerated protons are of 
about 200 MeV (cf. Tab.~1), which corresponds to $\beta^2\approx0.3$. Consequently, the truncation of the
interaction up to $\mathcal{O}(\beta^2)$ is justified and higher-order contributions such as those up to 
$\mathcal{O}(\beta^4)$ in Ref.~\cite{Jaen} will not be taken into account.

Equation (\ref{motion_coul}) is a differential algebraic equation of the form
\begin{eqnarray}
\frac{d\bm{\beta}_1}{dt}&=&f_1\left(\bm{r}_1,\hdots,\bm{r}_N,\bm{\beta}_1,\hdots,\bm{\beta}_N,\frac{d\bm{\beta}_1}{dt},\hdots,\frac{d\bm{\beta}_N}{dt},t\right),\nonumber\\
&\vdots&\nonumber\\
\frac{d\bm{\beta}_N}{dt}&=&f_N\left(\bm{r}_1,\hdots,\bm{r}_N,\bm{\beta}_1,\hdots,\bm{\beta}_N,\frac{d\bm{\beta}_1}{dt},\hdots,\frac{d\bm{\beta}_N}{dt},t\right).
\end{eqnarray}
For each time step the system
has to be solved algebraicly with respect to $\frac{d\bm{\beta}_1}{dt},\hdots,\frac{d\bm{\beta}_N}{dt}$. Then,
a standard fourth-order Runge-Kutta algorithm is used to integrate the remaining 
system of ordinary differential equations.

\begin{acknowledgments}

ZH acknowledges conversations with Y.~I.~Salamin. Supported by Helmholtz Alliance HA216/EMMI.

\end{acknowledgments}

%\bibliography{lit2}

\begin{thebibliography}{10}
\expandafter\ifx\csname url\endcsname\relax
  \def\url#1{\texttt{#1}}\fi
\expandafter\ifx\csname urlprefix\endcsname\relax\def\urlprefix{URL }\fi
\providecommand{\bibinfo}[2]{#2}
\providecommand{\eprint}[2][]{\url{#2}}

\bibitem{med1}
\bibinfo{author}{{Yock, T. I. \& Tarbell, N. J.}}
\newblock \bibinfo{title}{{Technology Insight: proton beam radiotherapy for
  treatment in pediatric brain tumors}}.
\newblock \emph{\bibinfo{journal}{{Nature Clinical Practice Oncology}}}
  \textbf{\bibinfo{volume}{1}}, \bibinfo{pages}{97--103}
  (\bibinfo{year}{2004}).

\bibitem{med2}
\bibinfo{author}{{Levin, M. P., Kooy, H., Loeffler, J. S. \& DeLaney, T. F.}}
\newblock \bibinfo{title}{Proton beam therapy}.
\newblock \emph{\bibinfo{journal}{{British Journal of Cancer}}}
  \textbf{\bibinfo{volume}{93}}, \bibinfo{pages}{849--854}
  (\bibinfo{year}{2005}).

\bibitem{debus1}
\bibinfo{author}{{J\"akel, O., Kr\"amer, M., Karger, C. P. \& Debus, J.}}
\newblock \bibinfo{title}{Treatment planning for heavy ion radiotherapy:
  clinical implementation and application}.
\newblock \emph{\bibinfo{journal}{Phys. Med. Biol.}}
  \textbf{\bibinfo{volume}{46}}, \bibinfo{pages}{1101--1116}
  (\bibinfo{year}{2001}).

\bibitem{cancer}
\bibinfo{author}{{Combs, S. E. {\it et al.}}}
\newblock \bibinfo{title}{Carbon ion radiotherapy for pediatric patients and
  young adults treated for tumors of the skull base}.
\newblock \emph{\bibinfo{journal}{Cancer}} \textbf{\bibinfo{volume}{115}},
  \bibinfo{pages}{1348--1355} (\bibinfo{year}{2009}).

\bibitem{lithog}
\bibinfo{author}{{van Kan, J. A., Bettiol, A. A. \& Watt, F.}}
\newblock \bibinfo{title}{{Three-dimensional nanolithography using proton beam
  writing}}.
\newblock \emph{\bibinfo{journal}{Appl. Phys. Lett.}}
  \textbf{\bibinfo{volume}{83}}, \bibinfo{pages}{1629--1631}
  (\bibinfo{year}{2003}).

\bibitem{ledingham}
\bibinfo{author}{{Ledingham, K. W. D., McKenna, P. \& Singhal, R. P.}}
\newblock \bibinfo{title}{{Applications for nuclear phenomena generated by
  ultra-intense lasers}}.
\newblock \emph{\bibinfo{journal}{Science}} \textbf{\bibinfo{volume}{300}},
  \bibinfo{pages}{1107--1111} (\bibinfo{year}{2003}).

\bibitem{lhc}
\bibinfo{title}{{LHC--The Large Hadron Collider}}.
\newblock \bibinfo{note}{\url{http://lhc.web.cern.ch/lhc/}}.

\bibitem{eli}
\bibinfo{title}{{The Extreme Light Infrastructure European Project (ELI).
  Scientific Case}} (\bibinfo{year}{2007}).
\newblock
  \bibinfo{note}{\url{http://www.extreme-light-infrastructure.eu/pictures/ELI-%
scientific-case-id17.pdf}}.

\bibitem{hiper}
\bibinfo{title}{{High Power Laser Energy Research (HiPER). HiPER technical
  background and conceptual design report}}.
\newblock
  \bibinfo{note}{\url{http://www.hiperlaser.org/docs/tdr/HiPERTDR2.pdf}}.

\bibitem{dunne-pers}
\bibinfo{author}{{Dunne, M.}}
\newblock \bibinfo{title}{{Laser-driven particle accelerators}}.
\newblock \emph{\bibinfo{journal}{Science}} \textbf{\bibinfo{volume}{312}},
  \bibinfo{pages}{374--376} (\bibinfo{year}{2006}).

\bibitem{plasma-el}
\bibinfo{author}{{Malka, V. {\it et al.}}}
\newblock \bibinfo{title}{{Electron acceleration by a wake field forced by an
  intense ultrashort laser pulse}}.
\newblock \emph{\bibinfo{journal}{Science}} \textbf{\bibinfo{volume}{298}},
  \bibinfo{pages}{1596--1600} (\bibinfo{year}{2002}).

\bibitem{plasma1}
\bibinfo{author}{{Hegelich, B. M. {\it et al.}}}
\newblock \bibinfo{title}{{Laser acceleration of quasi-monoenergetic MeV ion
  beams}}.
\newblock \emph{\bibinfo{journal}{Nature}} \textbf{\bibinfo{volume}{439}},
  \bibinfo{pages}{441--444} (\bibinfo{year}{2006}).

\bibitem{plasma2}
\bibinfo{author}{{Schwoerer, H. {\it et al.}}}
\newblock \bibinfo{title}{{Laser-plasma acceleration of quasi-monoenergetic
  protons from microstructured targets}}.
\newblock \emph{\bibinfo{journal}{Nature}} \textbf{\bibinfo{volume}{439}},
  \bibinfo{pages}{445--448} (\bibinfo{year}{2006}).

\bibitem{plasma3}
\bibinfo{author}{{Fuchs, J. {\it et al.}}}
\newblock \bibinfo{title}{{Laser-driven proton scaling laws and new paths
  towards energy increase}}.
\newblock \emph{\bibinfo{journal}{Nature Physics}}
  \textbf{\bibinfo{volume}{2}}, \bibinfo{pages}{48--54} (\bibinfo{year}{2006}).

\bibitem{plasma4}
\bibinfo{author}{{Robson, L. {\it et al.}}}
\newblock \bibinfo{title}{{Scaling of proton acceleration driven by
  petawatt-laser-plasma interactions}}.
\newblock \emph{\bibinfo{journal}{Nature Physics}}
  \textbf{\bibinfo{volume}{3}}, \bibinfo{pages}{58--62} (\bibinfo{year}{2007}).

\bibitem{plasma5}
\bibinfo{author}{{Maksimchuk, A., Gu, S., Flippo, K., Umstadter, D. \&
  Bychenkov, V. Y.}}
\newblock \bibinfo{title}{{Forward ion ccceleration in thin films driven by a
  high-intensity laser}}.
\newblock \emph{\bibinfo{journal}{Phys. Rev. Lett.}}
  \textbf{\bibinfo{volume}{84}}, \bibinfo{pages}{4108--4111}
  (\bibinfo{year}{2000}).

\bibitem{plasma6}
\bibinfo{author}{{Snavely, R. A. {\it et al.}}}
\newblock \bibinfo{title}{{Intense high-energy proton beams from petawatt-laser
  irradiation of solids}}.
\newblock \emph{\bibinfo{journal}{Phys. Rev. Lett.}}
  \textbf{\bibinfo{volume}{85}}, \bibinfo{pages}{2945--2948}
  (\bibinfo{year}{2000}).

\bibitem{plasma7}
\bibinfo{author}{{Karsch, S. {\it et al.}}}
\newblock \bibinfo{title}{{High-intensity laser induced ion acceleration from
  heavy-water droplets}}.
\newblock \emph{\bibinfo{journal}{Phys. Rev. Lett.}}
  \textbf{\bibinfo{volume}{91}}, \bibinfo{pages}{015001}
  (\bibinfo{year}{2003}).

\bibitem{plasma8}
\bibinfo{author}{{Romagnani, L. {\it et al.}}}
\newblock \bibinfo{title}{{Dynamics of electric fields driving the laser
  acceleration of multi-MeV protons}}.
\newblock \emph{\bibinfo{journal}{Phys. Rev. Lett.}}
  \textbf{\bibinfo{volume}{95}}, \bibinfo{pages}{195001}
  (\bibinfo{year}{2005}).

\bibitem{plasma9}
\bibinfo{author}{{Cowan, T. E. {\it et al.}}}
\newblock \bibinfo{title}{{Ultra-low emittance, high current proton beams
  produced with a laser-virtual cathode sheath accelerator}}.
\newblock \emph{\bibinfo{journal}{Nucl. Instr. Meth. Phys. Res.}}
  \textbf{\bibinfo{volume}{544}}, \bibinfo{pages}{277--284}
  (\bibinfo{year}{2005}).

\bibitem{plasma10}
\bibinfo{author}{{Albright, B. J. {\it et al.}}}
\newblock \bibinfo{title}{{Theory of laser acceleration of light-ion beams from
  interaction of ultrahigh-intensity lasers with layered targets}}.
\newblock \emph{\bibinfo{journal}{Phys. Rev. Lett.}}
  \textbf{\bibinfo{volume}{97}}, \bibinfo{pages}{115002}
  (\bibinfo{year}{2006}).

\bibitem{plasma11}
\bibinfo{author}{{Tajima, T. \& Dawson, J. M.}}
\newblock \bibinfo{title}{{Laser electron accelerator}}.
\newblock \emph{\bibinfo{journal}{Phys. Rev. Lett.}}
  \textbf{\bibinfo{volume}{43}}, \bibinfo{pages}{267--270}
  (\bibinfo{year}{1979}).

\bibitem{willi}
\bibinfo{author}{Mackinnon, A.~J.} \emph{et~al.}
\newblock \bibinfo{title}{{Effect of plasma scale length on multi-MeV proton
  production by intense laser pulses}}.
\newblock \emph{\bibinfo{journal}{Phys. Rev. Lett.}}
  \textbf{\bibinfo{volume}{86}}, \bibinfo{pages}{1769--1772}
  (\bibinfo{year}{2001}).

\bibitem{badziak}
\bibinfo{author}{{Badziak, J.}}
\newblock \bibinfo{title}{{Laser-driven generation of fast particles}}.
\newblock \emph{\bibinfo{journal}{Opto-Electr. Review}}
  \textbf{\bibinfo{volume}{15}}, \bibinfo{pages}{1--12} (\bibinfo{year}{2007}).

\bibitem{haaland}
\bibinfo{author}{{Haaland, C. M.}}
\newblock \bibinfo{title}{{Laser electron acceleration in vacuum}}.
\newblock \emph{\bibinfo{journal}{Opt. Commun.}}
  \textbf{\bibinfo{volume}{114}}, \bibinfo{pages}{280--284}
  (\bibinfo{year}{1995}).

\bibitem{esarey1}
\bibinfo{author}{{Esarey, E., Sprangle, P. \& Krall, J.}}
\newblock \bibinfo{title}{{Laser acceleration of electrons in vacuum}}.
\newblock \emph{\bibinfo{journal}{Phys. Rev. E}} \textbf{\bibinfo{volume}{52}},
  \bibinfo{pages}{5443--5453} (\bibinfo{year}{1995}).

\bibitem{sal00}
\bibinfo{author}{{Salamin, Y. I., Keitel, C. H.}}
\newblock \bibinfo{title}{{Subcycle high electron acceleration by crossed laser
  beams}}.
\newblock \emph{\bibinfo{journal}{Appl. Phys. Lett.}}
  \textbf{\bibinfo{volume}{77}}, \bibinfo{pages}{1082--1084}
  (\bibinfo{year}{2000}).

\bibitem{pfs}
\bibinfo{author}{{Major, Z. {\it et al.}}}
\newblock \bibinfo{title}{Basic concepts and current status of the petawatt
  field synthesizer – a new approach to ultrahigh field generation}.
\newblock \emph{\bibinfo{journal}{The Review of Laser Engineering}}
  \textbf{\bibinfo{volume}{37}}, \bibinfo{pages}{431--436}
  (\bibinfo{year}{2009}).

\bibitem{bahk}
\bibinfo{author}{{Bahk, S.-W. {\it et al.}}}
\newblock \emph{\bibinfo{journal}{Opt. Lett.}} \textbf{\bibinfo{volume}{29}},
  \bibinfo{pages}{2837--2839} (\bibinfo{year}{2004}).

\bibitem{sal-apb}
\bibinfo{author}{{Salamin, Y. I.}}
\newblock \bibinfo{title}{{Fields of a Gaussian beam beyond paraxial
  approximation}}.
\newblock \emph{\bibinfo{journal}{Appl. Phys. B}}
  \textbf{\bibinfo{volume}{86}}, \bibinfo{pages}{319--326}
  (\bibinfo{year}{2007}).

\bibitem{sal-prl2}
\bibinfo{author}{{Salamin, Y. I., Harman, Z. \& Keitel, C. H.}}
\newblock \bibinfo{title}{{Direct high-power laser acceleration of ions for
  medical applications}}.
\newblock \emph{\bibinfo{journal}{Phys. Rev. Lett.}}
  \textbf{\bibinfo{volume}{100}}, \bibinfo{pages}{155004}
  (\bibinfo{year}{2008}).

\bibitem{scheid}
\bibinfo{author}{{Wang, J. X. {\it et al.}}}
\newblock \bibinfo{title}{High-intensity laser-induced electron acceleration in
  vacuum}.
\newblock \emph{\bibinfo{journal}{Phys. Rev. E}} \textbf{\bibinfo{volume}{60}},
  \bibinfo{pages}{7473--7478} (\bibinfo{year}{1999}).

\bibitem{space-temp}
\bibinfo{author}{{Yan, Z. {\it et al.}}}
\newblock \bibinfo{title}{{Accurate description of ultra-short tightly focused
  Gaussian laser pulses and vacuum laser acceleration}}.
\newblock \emph{\bibinfo{journal}{Appl. Phys. B}}
  \textbf{\bibinfo{volume}{81}}, \bibinfo{pages}{813--819}
  (\bibinfo{year}{2005}).

\bibitem{zolotorev}
\bibinfo{author}{Stupakov, G.~V.} \& \bibinfo{author}{Zolotorev, M.~S.}
\newblock \bibinfo{title}{Ponderomotive laser acceleration and focusing in
  vacuum for generation of attosecond electron bunches}.
\newblock \emph{\bibinfo{journal}{Phys. Rev. Lett.}}
  \textbf{\bibinfo{volume}{86}}, \bibinfo{pages}{5274--5277}
  (\bibinfo{year}{2001}).

\bibitem{eickhoff}
\bibinfo{author}{{Eickhoff, H. {\it et al.}}}
\newblock \bibinfo{title}{{HIT--Heidelberg Ion beam Therapy. Scientific Case.}}
\newblock \bibinfo{note}{\url{http://www-aix.gsi.de/~spiller/facilit_ep00.ps}}.

\bibitem{hit}
\bibinfo{title}{{HIT--Heidelberg Ion beam Therapy. Facts in short.}}
\newblock
  \bibinfo{note}{\url{http://www.klinikum.uni-heidelberg.de/HIT-Facts-in-short%
.117995.0.html?&L=en}}.

\bibitem{eli-costs}
\bibinfo{author}{{Gerstner, E.}}
\newblock \bibinfo{title}{{Laser physics: extreme light}}.
\newblock \emph{\bibinfo{journal}{Nature}} \textbf{\bibinfo{volume}{446}},
  \bibinfo{pages}{16--18} (\bibinfo{year}{2007}).

\bibitem{jackson}
\bibinfo{author}{{Jackson, J. D.}}
\newblock \emph{\bibinfo{title}{{Classical Electrodynamics}}}
  (\bibinfo{publisher}{John Wiley \& Sons}, \bibinfo{year}{1999}),
  \bibinfo{edition}{third} edn.

\bibitem{Jaen}
\bibinfo{author}{Ja\'en, X.}, \bibinfo{author}{Llosa, J.} \&
  \bibinfo{author}{Molina, A.}
\newblock \bibinfo{title}{{A reduction of order two for infinite-order
  Lagrangians}}.
\newblock \emph{\bibinfo{journal}{Phys. Rev. D}} \textbf{\bibinfo{volume}{34}},
  \bibinfo{pages}{2302--2311} (\bibinfo{year}{1986}).

\end{thebibliography}
%\bibliographystyle{naturemag}
% \begin{thebibliography}{34}

% 
% \end{thebibliography}

\end{document}